\documentclass[sigconf]{acmart}

\AtBeginDocument{%
  \providecommand\BibTeX{{%
    \normalfont B\kern-0.5em{\scshape i\kern-0.25em b}\kern-0.8em\TeX}}}

\acmConference[HCIxB Workshop, CHI 2025]{ACM Conference on Human Factors in Computing Systems}{April 26 - May 1}{Yokohama, Japan}
\graphicspath{{./images/}}

\begin{document}

\title{Bridging Data Gaps and Building Knowledge Networks in Indian Football Analytics}

\author{Sneha Nanavati}
\email{sneha.nanavati@research.iiit.ac.in}
\affiliation{%
  \institution{International Institute of Information Technology, Hyderabad}
  \country{India}
}

\author{Nimmi Rangaswamy}
\email{nimmi.rangaswamy@iiit.ac.in}
\affiliation{%
  \institution{International Institute of Information Technology, Hyderabad}
  \country{India}
}


\begin{abstract}
The global rise of football analytics has rapidly transformed how clubs make strategic decisions~\cite{morgulev2018sports}. However, in India, the adoption of analytics remains constrained by institutional resistance, infrastructural limitations, and cultural barriers—challenges that grassroots innovation and low-cost data solutions have the potential to overcome. Despite the increasing popularity of the Indian Super League~\cite{scroll2020isl,chatterjee2021isl}, resource scarcity and fragmented governance continue to hinder the widespread adoption and impact of analytics. This mixed-methods study explores how informal, decentralized analytics communities—comprising amateur analysts and Twitter-based “data sleuths”—navigate these constraints through peer mentorship and grassroots innovation. Drawing on extensive digital ethnography, participant observation, and interviews, the study illustrates how these informal networks mitigate data scarcity, limited digital infrastructure, and institutional indifference while fostering skill development and professional growth.
Building on these insights, the paper proposes HCI interventions such as decentralized knowledge platforms to facilitate structured, cross-border peer mentorship and low-cost data solutions—including AI-assisted player tracking and mobile analytics dashboards—rooted in principles of frugal innovation~\cite{shauloff2023ai}. These interventions aim to bridge the data divide, support inclusive technical engagement in sport, and enhance analytics-driven decision-making in resource-constrained environments. This paper contributes to HCIxB’s focus on cross-border collaboration by highlighting how community-driven technological adaptation in the Global South can foster meaningful participation, skill-building, and long-term sustainability through informal learning networks and scalable, context-sensitive tools~\cite{vashistha2017cve,nawaz2021farming}.

\end{abstract}



\maketitle

\section{Introduction}
Data analytics has enabled football teams globally to challenge traditional powerhouses through strategic recruitment, performance analysis, and tactical innovations—revolutionising decision-making practices in the game. Billy Beane’s seminal Moneyball method in baseball is now extensively adopted across all sports, including football, basketball, cricket and even Formula 1. The posterchild for this revolution is clubs like Brentford FC’s Cinderella story of rising to the English Premier League within a few short seasons, built on sustainable growth driven by data, or the David vs Goliath tale of FC Midtjylland, a small Danish club propelled to global stardom through data-driven recruitment and strategy~\cite{robinson2021moneyball, midtjylland2020, lewis2003moneyball}.

While the global trend towards analytics-driven football continues to grow in Western contexts, Indian football remains plagued by institutional barriers, infrastructure inadequacies, and cultural resistance to data adoption. The launch of the Indian Super League (ISL) in 2014 marked increased nationwide viewership and fan engagement. ISL aimed to capitalise on India’s passion for football, particularly among fans of the English Premier League and European tournaments like the UEFA Champions League. Inspired by the IPL franchise model, the ISL adopted a similar structure to attract international coaches and players. By the 2019–2020 season, ISL was garnering 168 million viewers with 261 million impressions, and the following season saw a 16\% increase in pan-India viewership~\cite{chatterjee2021isl, scroll2020isl}. A 2018 survey found that 45\% of India’s urban population was interested in football, up from 30\% in 2013~\cite{scroll2020isl}. Hosting the 2017 FIFA Under-17 World Cup—drawing 1.2 million stadium spectators—was a landmark moment for Indian football. Youth development programs and club-level collaborations, such as those by Kerala Blasters FC and Hyderabad FC, have sought to foster domestic talent~\cite{rajiv2022reliance}.

Despite these advancements, Indian football continues to face significant challenges in achieving international success. The national team’s current FIFA ranking of 125th (as of October 2024) starkly contrasts with Asian peers like Japan and South Korea, ranked 15th and 22nd respectively~\cite{olympics2025}. Key limitations include lack of access to digital and physical infrastructure, inconsistent government support, and underdeveloped grassroots systems. A preference for intuition-based decision-making and resistance from coaches and scouts towards analytics continues to hamper innovation.

This study explores how a decentralised football analytics community in India—comprising self-taught “data sleuths” active on social media, formally employed club analysts, and football journalists—navigates these barriers. Though diverse in role and background, these actors collectively form an informal, evolving ecosystem centred on tactical analysis, data storytelling, and open knowledge exchange.

Using a mixed-method ethnographic approach, this research investigates the community of self-taught data analysts in the Indian football scene. Their ground-up network thrives on informal learning and mentorship, overcoming computational limitations to foster frugal innovation. It identifies opportunities for HCI interventions such as decentralised knowledge platforms and low-cost, frugal technologies to sustainably improve analytics capacity and digital inclusion within football’s resource-constrained settings. This paper builds on the impact of community-driven innovation and highlights the potential of informal learning networks and scalable technological interventions in the Global South.

\section{Related Work}

This paper builds on prior research in HCI4D and ICTD that explores how informal learning infrastructures emerge and thrive in resource-constrained environments. Vashistha et al.’s work on community-led video education (CVE) introduces a model for disseminating digital financial knowledge among low-income, low-literacy users in rural India~\cite{vashistha2017cve}. Reflecting the CVE model’s peer-driven approach, the Indian football analytics community similarly leverages Twitter and direct messages as informal mentoring channels and live feedback loops. Nawaz et al. showed how local farming communities in Punjab depend on peer-trusted networks to interpret agricultural knowledge, underscoring the significance of community-driven knowledge sharing~\cite{nawaz2021farming}. Likewise, Rony et al. observed that low-income Bangladeshi women adopted digital financial services through peer support and hands-on use, overcoming the limitations of poorly designed digital interfaces~\cite{rony2021dfs}. Like these Bangladeshi working women, the analysts—despite initial unfamiliarity—developed confidence through peer support.

In line with this body of work, this paper examines how Indian football analysts use informal networks and open-source tools to create functional learning pathways despite the absence of institutional support.

\section{Methodology}

This study employs an iterative-inductive qualitative ethnographic approach to explore how the football analytics community in India functions comprehensively. Ethnography was chosen to investigate the nuances interplaced between technical, social and cultural dynamics among the diverse participants of football analytics. This ranged from amateur data enthusiasts to professional analysts and sports journalists. 

\subsection{\textbf{Research Design}}

The iterative-inductive research framework for ethnographic analysis was designed based on the methodological guidelines by O’Reilly and by Scott-Jones and Watt~\cite{oreilly2012ethnographic,scottjones2010ethnography}. This approach allowed for the cyclical integration of data collection and analysis, ensuring thoroughness and flexibility. Triangulation of interviews, participant observations and literature review provided rich, multi-dimensional insights into participants' experiences, practises and perceptions.

\subsection{\textbf{Participant Profiles and Selection}}

Participants were identified and categorised into distinct interconnected groups based on their roles and involvement within the Indian football analytics community: 

\begin{itemize}
    \item \textbf{Data Sleuths}: Amateur analysts are mainly active on Twitter (now X), informally experimenting with data extraction and analysis. They are highly self-driven and engage in informal mentorship, peer collaboration, skill development, and knowledge sharing. They are often fans of a team who start analytics as a way to better understand their team’s gameplay and pivot into learning new skills and showcasing them online. Many of these individuals aspire and successfully transition into professional sports analytics roles in football teams. These individuals are heavily dependent on open-source tools and freely available data. 
    \item \textbf{\textbf{Football Club Data Analysts: }}They are formally employed or contracted by football clubs and have direct involvement in recruitment, match preparation and tactical analysis. Many analysts begin their careers within the online spaces before transitioning into formal professional roles. 
roles.\textbf{}
    \item \textbf{Football Writers and Data Journalists}: These writers merge analytics with journalism, publishing insightful data-driven narratives for sports media and digital publication. A need for more data-centred stories is also coming from the mainstream audience. Their work contributes to a broader public understanding of football analytics. 
\end{itemize}

\subsection{\textbf{Data Collection Methods}}

Data collection was conducted through multiple structures and unstructured methods tailored to the participant category:

\subsubsection{\textbf{Structured Interviews}}
Interviews conducted via video and audio calls or written formats collected details on participants' backgrounds, analytics practice and personal experiences. Specific questions were asked based on the participants’s unique sights. For example, Data Sleuths were asked how they started in the data analytics field and what kept them engaged. Similarly, professional analysts were asked about the tools and methods that impact their club’s tactical strategies most. 

\subsubsection{\textbf{Unstructured Methods}}
\begin{itemize}
    \item \textbf{\textbf{Digital Ethnography and Participant Observation: }}This study included observational fieldwork on platforms such as Twitter (X), forums and The AwayEnd, which allowed close observations of real-time interactions and learning processes, including spontaneous collaborative activities. Participating in these spaces offered a glimpse of community dynamics, strategies members use, and challenges they face. 
\textbf{}
    \item \textbf{\textbf{Field Notes and Documentary Analysis: }}Detailed observation notes, interview transcripts and documentary sources such as blogs, articles and tweets add to the context of the ethnographic dataset, especially in fleshing out the analytical narratives emerging from structured interviews.
\end{itemize}

Participants were identified and recruited through their digital activities in analytics spaces, leveraging familiarity within the online analytics community. 

\subsection{\textbf{Data Analysis}}

The analysis involved a rigorous iterative-inductive coding process:

\subsubsection{\textbf{Data Organisation}}

Interview recordings (audio and video) were transcribed, categorized, and organized by participant profiles, while field notes from observations and insights from the literature review were systematically integrated into the analysis.

\subsubsection{\textbf{Coding Process}}

Coding began with an initial review of interview transcripts, highlighting segments relevant to the research interests, such as mentorship pathways, infrastructural challenges, analytics practices and data disparities. Recurring patterns such as "community engagement," "data scarcity challenges," and "informal analytics training” emerged. Revisiting and refining the coded data ensured thematic coherence and robustness. This included comparing interview segments across participant categories, varying thematic consistency and analytical clarity. 

\subsubsection{\textbf{Theme Development}}

The coding process outlined broader thematic clusters, demonstrating key socio-technical dynamics within Indian football analytics. Some of the prominent themes include: 

\begin{itemize}
    \item \textbf{\textbf{Community Engagement and Informal Mentorship: }}Highlighting how informal interactions facilitate skill development and career progression.\textbf{}
    \item \textbf{\textbf{Impact of Analytics on Club Performance: }}Identifying tangible examples of analytics-driven strategies improving team tactics and player recruitment despite limited resources. It also explores the real-life challenges of implementing data tools among other stakeholders. 
    \item \textbf{\textbf{Institutional and Infrastructual Limitations: }}Exposing constant challenges ranging from fragmented data and low-tech infrastructure to institutional resistance towards analytics adoption. \textbf{}
\end{itemize}
These themes were evaluated for their relevance to the research objective, their recurrence across multiple interactions, and their contribution to understanding the socio-cultural dynamics of analytics and football infrastructure.

\subsection{\textbf{Ethical Considerations}}

Ethical practises were prioritised by focusing on participant anonymity, confidentiality and informed consent. The participants were informed about the study's purpose and use and were offered the option to withdraw at any time. The details of participants and their workplaces were anonymised to ensure confidentiality and trustworthiness. 

\subsection{\textbf{Challenges and Limitations}}

This study acknowledges certain constraints, such as variance in participant engagement and response methods. Not all targeted participants actively responded, reflecting real-world research challenges involving informal groups. Despite a focus towards diversity, gender representation remained skewed towards male candidates, which could potentially impact the absence of certain trends and themes. 

\section{Findings}

This study presents four key themes that emerged through iterative ethnographic analysis. Each cluster reveals how individuals in Indian football analytics communities adapt to and rework infrastructural constraints, institutional resistance, and the absence of formal pathways for learning and employment.

\subsection{\textbf{Informal Learning and Community-Driven Knowledge Creation}}

Across all participant profiles, informal learning surfaced as a central mode of skill acquisition. Analysts often entered the space with little to no formal training in data science or football-specific analytics. They developed their expertise through freely available online resources, datasets, libraries, peer contributions, and informal mentorship networks.

\begin{quote}
\textit{“People with no coding background got into it due to their interest in football, and people already knowledgeable helped to drive the movement.”} – \textbf{H.}
\end{quote}

\begin{quote}
\textit{“The online community was crucial (in my growth). Everyone had a niche—one person was into Tableau, another into Python or scraping. You’d just ask (on Twitter), and someone would help.”} – \textbf{S}
\end{quote}

Social media—particularly Twitter—emerged as a living infrastructure for mentorship and skill-sharing, creating space for collaboration and distribution. Analysts described it as a porous, decentralised classroom:

\begin{quote}
\textit{“It was almost like a public lab. You would post your work, someone would DM you corrections or reply with a share a better approach. That kind of feedback loop was great for a beginner like me.”} – \textbf{M}
\end{quote}

This kind of decentralised pedagogy filled the gap left by the absence of formal football analytics education, forming the bottom-up infrastructure of expertise. These peer-driven networks form the core of what some participants called a “community of practice”, where the focus is on learning and passing their knowledge to other up-and-coming data sleuths. 

\subsection{\textbf{Institutional Friction: Influence, Legibility, and Translating Analytics}}

Along with the growing opportunities in professional football environments, analysts encounter the challenge of gaining trust and institutional support. Coaches, club directors and technical staff may place analytics on the periphery of decision-making. Talking about the challenge of getting everyone on board, one analyst observes:

\begin{quote}
\textit{“\textit{I mostly communicate with the director—not the coach. The coach isn’t data-oriented. My findings might get passed on, but there’s no guarantee they’ll be used.}”} – \textbf{S}
\end{quote}

Participants noted that data accuracy was not the only challenge—they also faced hurdles in translating insights into actionable terms. Making data legible and tactically meaningful for coaches accustomed to intuitive decision-making required navigating a steep learning curve.

\begin{quote}
\textit{“Even though you can use clustering or visualisations, it’s all about how the coach sees it. It’s important for the players and coaches to keep things easy to read and understand. Otherwise it (insights) won’t get used.”} – \textbf{N}
\end{quote}

Institutional support can also be precarious. One of the data analysts noted that changes in leadership, such as appointing a  new coach, could invalidate their entire workflow  models and frameworks built over several seasons: 

\begin{quote}
\textit{“When the head coach changed, my whole model became irrelevant. He did not believe in the data and dismissed it. You might have to start from scratch depending on the new coach’s preferences.”} – \textbf{V}
\end{quote}

The lived experiences of analysts reveal a central tension in analytics work within ISL clubs: balancing technical rigour with interpretive flexibility in the face of volatility and cultural friction. In HCI, technical accuracy alone isn’t sufficient—data tools must also be understandable and adaptable to remain effective in shifting contexts. This calls for designing with translatability and resilience in mind.

\subsection{\textbf{Frugal Innovation: Making Analytics Work Without Infrastructure}}

A recurring theme across interviews was the creative adaptation of low-cost tools and public datasets to circumvent infrastructural deficiencies. Participants consistently emphasised their reliance on open platforms like \textbf{FBref, WhoScored, Transfermarkt}, or basic \textbf{GPS tracker }data paired with self-written scraping scripts and libraries. 

\begin{quote}
\textit{“We can’t afford expensive tracking systems. So we use VO cameras, scrape public data, and try to extract whatever positional metrics we can. It’s not ideal, but it’s something.”} – \textbf{K}
\end{quote}

For many participants, the distinguishing factor was not using elite tools but seeing how they could effectively work with limited data—what one described as “getting insights from scraps.”

\begin{quote}
\textit{“You’re stitching together partial datasets, making tactical visualisations from YouTube clips or event data. You learn to prioritise clarity over perfection.”} – \textbf{H}
\end{quote}

One analyst spoke about repurposing visualisation tools and statistical packages designed for other sports for football: 
\begin{quote}
\textit{“\textit{I used k-clustering in a really basic way—just subtracting and comparing metrics—to group similar players. Nothing fancy. But that alone made our scouting feel more systematic.}”} – \textbf{M}
\end{quote}

Rather than aspiring to create a high-fidelity system comparable to their European counterparts, analysts present their approach as frugal, focused on clarity, functionality, and explicable to decision-makers. 

\subsection{\textbf{Social Media as Infrastructure for Learning and Hiring}}

This digital community serves as a learning environment and gradually evolves into a springboard for formal roles in data analytics.  Participants use their social media presence, especially content on Twitter, as a live portfolio. 

\begin{quote}
\textit{“I started with posting charts on Twitter. People noticed. Eventually, clubs reached out. It was never formal recruitment—just visibility.”} – \textbf{S}
\end{quote}

Another participant provided examples of peers who had followed similar trajectories:

\begin{quote}
\textit{“People have been hired based on their work in community spaces. MT got hired by Barnsley, and AR joined Dundee after years of online posts and blogs.”} – \textbf{H}
\end{quote}

This visibility is vital, especially in the absence of a formal qualification in football data analysis where standardised certification remains rare. 

\begin{quote}
\textit{“There’s no clear roadmap for becoming an analyst in India. Your work has to speak for itself—and online is the only place it can be seen.”} – \textbf{K}
\end{quote}

Digital spaces like Twitter serve as a skill incubator and employment pipeline. This pipeline from data sleuth to formal club role is a testament to community learning, underscoring the socio-economic potential of designing platforms that enhance visibility, validation, and learning.

\section{Discussion and HCI Implications}

The findings from this study highlight the resourceful practices of the footballing community in India while offering a textured glimpse of how knowledge infrastructure emerges, evolves, and sustains itself amidst institutional inertia and constraints. These analytics communities function as sites of learning, creating a bubble for career formation and technological improvisation, culminating in collective aspiration. They present a compelling case for how HCI research can engage with sports, data, and decentralised learning systems in the Global South. 

\subsection{\textbf{Digital Community as Infrastructure}}

The subjects of this study reveal how communities can operate as a substitute infrastructure by filling in for absent formal learning institutions. Twitter threads, open-source scripts, and DM-based mentorships on social media function as informal training grounds where analysts develop both technical skills and analytic thinking. These dynamics are reminiscent of video education in the financial literacy domain or WhatsApp-based agriculture extension networks. However, unlike these grassroots innovations that remain localised to a smaller group or geography, the football analytics community has shown itself to be translocated. They are networking with global actors while being rooted in local frictions. 
For the broader HCI community, the Indian football analytics scene offers a case study of how learning and technical skill-building can emerge and sustain itself without formal structures. Here, outputs like analysis and visualisations are closely tied to the social effort of teaching, sharing, and mentoring. These informally structured systems of knowledge production deserve more attention as legitimate, community-led pedagogies.

\subsection{\textbf{Translating Data into Trust}}

A recurring tension featured throughout this study from various participants is the friction between analytics and intuition. Their experiences highlighted the daily negotiations between objective knowledge and subject expertise. Analysts and their technical efforts must routinely translate and defend their statistical outputs to coaches and players into easily accessible narratives. While this data translation might be routine, it is a strategic act where analysts must constantly prove that their insights are credible, trustworthy, and compatible with the coach’s beliefs and ways of working. This can make using data not just a technical act but a political one, shaped by authority, persuasion and culture.  

For HCI, the challenge is improving data accuracy in low-resource settings while designing systems people can trust and use. In football, trust doesn’t come from precise results alone. Coaches, players, and club staff need to understand what the data means and how it connects to their on-ground decisions. This means analytics tools must allow for flexibility: outputs may need to be simplified, reframed, or adapted to fit different coaching styles and communication preferences. Designing for legibility, not just functionality, becomes key.

\subsection{\textbf{Frugal Innovation and Situated Tooling}}

The analysts in this study are aware of their resource limitations and are not seeking a replication of elite European infrastructures instead, they are building good enough systems from publicly available data, scraping scripts, and open-source libraries. These actions should not be seen only as deficiencies but as design chives rooted in realities and personal goals. The goal is persuasion, not perfection. 

This redirects the focus of the HCI and CSCW community to the impact of frugality and repair as innovation logic. While designing for these communities, it is important to prioritise modularity, rapid feedback and visual simplicity over feature-rich rigid solutions.

\subsection{\textbf{What Remains Unseen}}

Despite its many vibrant participants, the Indian football analytics space observed in this study describes a narrow demographic scope. None of the primary analysts interviewed were women. Castle, class and language were rarely made explicit, but these subtly impacted access. For instance, fluency in English, access to devices with computing capacity, and internet bandwidth are all delimit factors based on who participants are and who gets seen. 

Extending these skills and resources to passionate individuals would require HCI to recognise what is present while engaging with the absences in the shadow. Solutions must be designed with and for the informal communities while being sensitive to their practices and omissions by engaging with individuals outside the urban, English-speaking, male-dominated spheres. Further exploration of how infrastructural interventions might redistribute visibility and participation remains to be studied. 

\subsection{\textbf{Proposed HCI Interventions}}

\subsubsection{\textbf{Decentralised Mentorship and Knowledge Platforms}}

Building structured open-access repositories based on the informal Twitter-to-club analyst pipeline. These include curated tutorials, anonymised case studies, peer feedback systems, and compute resources. Such platforms would legitimise informal learning while making entry into this domain more streamlined for newcomers. Building these solutions alongside the community members is important, as well as keeping modularity, feedback, and ease of access front and centre. 

\subsubsection{\textbf{Frugal Toolkits for Data Access and Visualisation}}

Another proposed intervention is the development of a low-resource toolkit for data scripting, clustering, and visual storytelling. Analytics tools and data solutions should be designed to function offline, in low-bandwidth environments, and with mobile-first usability—requiring minimal computational power. Crucially, they should build upon existing practices rather than impose entirely new workflows. Indian football analytics presents a compelling case study of bottom-up innovation—where individual actors piece together influence, expertise, and career trajectories despite systemic constraints. For the HCI community, this shifts the focus from designing systems that scale to supporting systems that sustain—especially those fighting for legitimacy and visibility within a complex, evolving knowledge ecosystem.

\section{Conclusion}

This paper explores how analytics communities in Indian football navigate institutional resistance, infrastructural limitations, and cultural frictions to build a decentralized ecosystem for knowledge and skill development. Using an ethnographic approach, it examines how analysts operate outside formal structures, relying on online peer networks, open-source tools, and social media to learn, collaborate, and pursue professional opportunities. The study contributes to HCIxB in three key ways. First, it shows how informal learning communities function as support systems—providing training, feedback, and collaboration—where formal institutions are absent. It frames these efforts as infrastructural labour, recognizing the often-overlooked work of sustaining knowledge-sharing ecosystems.

Second, it highlights how analysts must continually translate their insights to make them accessible and persuasive to coaches, players, and decision-makers. This underscores the politics of data legibility—shifting attention from merely making data actionable to making it trustworthy and contextually meaningful.

Finally, the paper calls for HCI interventions that support and scale these grassroots efforts, strengthening the informal infrastructures that already exist.

Overall, this work urges the HCI community to pay closer attention to the social relationships, everyday routines, and practical challenges involved in deploying technology in resource-constrained settings. In this context, football is not just a sport or industry—it is a platform for building skills, gaining recognition, and imagining new futures. It emerges as a site of aspiration and reinvention.

\section{Limitations and Future Work}

This study focuses on a specific segment of the Indian football analytics community—primarily male, English-speaking participants from urban backgrounds. While this reflects the segment of the community that was most accessible during the research, it also means that many other perspectives—particularly those from less visible or marginalized groups—may be underrepresented. Future research should address these structural exclusions by exploring the experiences of women analysts, amplifying regional voices, and including those working outside dominant linguistic norms. Further work could also involve co-designing mentorship platforms and low-resource analytics toolkits in collaboration with the analyst community, emphasising evaluating their usability, adoption, and long-term sustainability.

\bibliographystyle{unsrt}
\bibliography{references}

\begin{thebibliography}{10}

\bibitem{morgulev2018sports}
Elia Morgulev, Ofer~H. Azar, and Ronnie Lidor.
\newblock Sports analytics and the big-data era.
\newblock {\em International Journal of Data Science and Analytics}, 5(4):213--222, 2018.

\bibitem{scroll2020isl}
Scroll Staff.
\newblock Indian football: Isl 2019-’20 records 51\% growth in viewership since start of season, says report.
\newblock \url{https://scroll.in/field/958030/indian-football-isl-2019-20-records-51-growth-in-viewership-since-start-of-season-says-report}, 2020.
\newblock Accessed April 10, 2025.

\bibitem{chatterjee2021isl}
Sayan Chatterjee.
\newblock Isl 2020-21 shows 16\% pan-india growth from last season’s viewership numbers.
\newblock \url{https://thebridge.in/football/isl-2020-21-shows-pan-india-growth-last-seasons-viewership/}, 2021.
\newblock Accessed April 10, 2025.

\bibitem{shauloff2023ai}
Dor Shauloff.
\newblock How ai helped put women’s football on the same footing as the men’s.
\newblock \url{https://pixellot.tv/use-cases/how-ai-helped-put-womens-football-on-the-same-footing-as-the-mens/}, 2023.
\newblock Accessed April 10, 2025.

\bibitem{vashistha2017cve}
Aditya Vashistha, Richard Anderson, and Rashmi Kanthi.
\newblock Community-led video education for digital financial services in india.
\newblock \url{https://www.adityavashistha.com/uploads/2/0/8/0/20800650/dfs_hcixb_2017.pdf}, 2017.
\newblock Accessed April 10, 2025.

\bibitem{nawaz2021farming}
Fareeda Nawaz, Abdul~Moeed Asad, and Maryam Mustafa.
\newblock “now i’m experienced and want to share”: Archiving and disseminating situated farming practices and knowledge, 2021.
\newblock Accepted at HCIxB 2021.

\bibitem{robinson2021moneyball}
Joshua Robinson.
\newblock A moneyball experiment in english soccer’s second tier.
\newblock \url{https://www.wsj.com/articles/barnsley-championship-promotion-moneyball-billy-beane-11621176691}, 2021.
\newblock Accessed April 10, 2025.

\bibitem{midtjylland2020}
The world’s most innovative club? the secrets behind liverpool’s champions league opponents midtjylland.
\newblock \url{https://www.goal.com/en/news/the-worlds-most-innovative-club-the-secrets-behind-liverpools-champions-league-opponents-midtjylland/1g6flveievyep1lywgw1c3g7d9}, 2020.
\newblock Accessed April 10, 2025.

\bibitem{lewis2003moneyball}
Michael Lewis.
\newblock {\em Moneyball}.
\newblock W. W. Norton \& Company, New York, 2003.

\bibitem{rajiv2022reliance}
Pranay Rajiv.
\newblock Reliance foundation young champs: Not just a feeder system for the isl, but a training academy with a difference.
\newblock \url{https://sportstar.thehindu.com/football/isl/reliance-foundation-youth-champs-isl-football-training-academy-neil-saunders/article66053379.ece}, 2022.
\newblock Accessed April 10, 2025.

\bibitem{olympics2025}
Indian football in asian games: The history, medals and results.
\newblock \url{https://www.olympics.com/en/news/indian-football-asian-games-results-medals}, 2025.
\newblock Accessed April 10, 2025.

\bibitem{rony2021dfs}
Rahat~Jahangir Rony, Syeda~Shabnam Khan, and Nova Ahmed.
\newblock “i didn’t understand but i was determined to learn”: Understanding the contrast of using dfs among the working women in bangladesh.
\newblock \url{https://drive.google.com/file/d/1ltJZMmjeNJa-3Ps1ADec38HGXuOwbDrR/view}, 2021.
\newblock Accessed April 10, 2025.

\bibitem{oreilly2012ethnographic}
Karen O'Reilly.
\newblock {\em Ethnographic Methods}.
\newblock Routledge, London, 2012.

\bibitem{scottjones2010ethnography}
Julie Scott-Jones and Sal Watt.
\newblock {\em Ethnography in Social Science Practice}.
\newblock Routledge, London, 2010.

\end{thebibliography}

\end{document}